\documentclass[article,10pt]{article}
\usepackage{graphicx,color, amsmath}
\usepackage{mathrsfs}
\usepackage{amsfonts, lscape}
\usepackage[english]{babel}
\usepackage[latin1]{inputenc}

\newcommand{\diff}{\mathrm{d}}

\ifx\pdfoutput\undefined
\usepackage[hypertex]{hyperref}
\else
\usepackage[pdftex,hypertexnames=false]{hyperref}
\fi

\begin{document}

\title{Dynamical System-based Robot Reaching Motions by Para-Model Control Approach \hspace{2cm} - A preliminary study -}
\author{{Lo\"ic Michel}}
\maketitle

\begin{quotation}                
\begin{center} {\bf Abstract} \end{center}
\noindent
In this report, we apply the proposed "para-model" framework in order to control the trajectory of a dynamical system-based robot. The optimization
of the dynamical performances in closed-loop is performed using a derivative-free optimization algorithm. 
\end{quotation}

\section{Model-free control approach}

The model-free control methodology has been originally proposed by Fliess $\&$ Join \cite{fliess2}, which is referred to as a self-tuning controller in \cite{kumar} and which has been widely and successfully 
applied to many mechanical and electrical processes. This control law has been designed to "robustify" {\it a priori} any "unknown" dynamical system for which not only uncertainties and 
unexpected modifications of the model parameter(s) are considered, but also switched models and models with time-delay(s)... 

\vspace{0.5cm}
The principle of this control law consists in building an ultra-local model of the controlled process from the measurements of the input and output signals, but the main 
disadvantage is that the derivative of the output signal is required. This ultra-local model is a part of an "auto-adjusting" or "extended" PI control and the performances 
are really good taking into account that no explicit model is {\it a priori} given - the control is only based on input $\&$ output signals.

One of the last contributions, called {\it para-model agent} (PMA) \cite{michel}, removes the use of the derivatives and replaces them by an 
initialization function. This contribution can be considered as a {\it derivative-free} $\&$ {\it model-free control scheme}. The last application, which has been successfully experimentally validated,
deals with the nonlinear control of the Epstein Frame, which is a device to characterize some physical properties of magnetic materials.

Based on the work of Khansari-Zadeh $\&$ Billard \cite{khansari}, we apply the proposed "para-model" framework in order to control the trajectory of a 
dynamical system-based robot, for which we aim to optimize the dynamical performances.

\newpage
\section{General Principle}

We consider a nonlinear SISO dynamical system to control:

\begin{equation}\label{eq:gen_sys}
u \mapsto y,  \quad \left\{ \begin{matrix}
\dot{x} = f_{nl}(x,u) \\
y = C x
\end{matrix} \right.
\end{equation}

\noindent
where $f_{nl}$ is a nonlinear system, the para-model agent is an application $(y^*, y) \mapsto u$ whose purpose is to control the output $y$ of (\ref{eq:gen_sys}) 
following an output reference $y^*$. In simulation, the system \ref{eq:gen_sys} is controlled in its "original formulation" without any modification / linearization.

\subsection{Definition of the closed-loop}

\begin{figure}[!h]
\centering
\includegraphics[width=11cm]{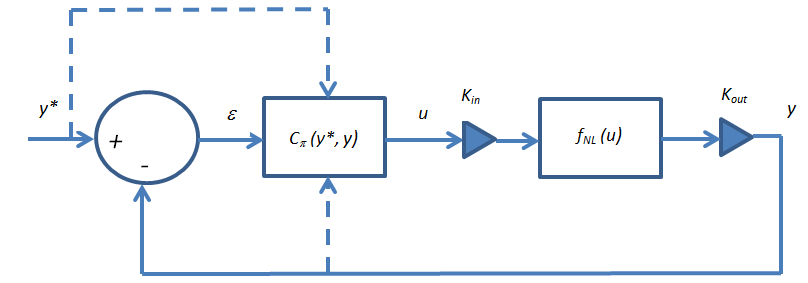}
\caption{Proposed PMA scheme to control a nonlinear system.}
\label{fig:CSM_gen}
\end{figure}

Consider the control scheme depicted in Fig. \ref{fig:CSM_gen} where $\mathcal{C}_{\pi}$ is the proposed PMA controller.

\subsection{Definition of the PMA algorithm}

For any discrete moment $t_k, \, k \in \mathbb{N}^*$, one defines the discrete controller $\mathcal{C}_{\pi}$ such that symbolically:
\begin{equation} \label{eq:iPI_discret_nm_eq}
\mathcal{C}_{\pi} : \begin{array}{c}
    \mathbb{R}^2 \rightarrow \mathbb{R}\\
    \displaystyle{  (y, y^*) \mapsto u_k =  \left. \int_0^t K_i \varepsilon_{k-1} d \, \tau \right|_{k-1} \underbrace{\left\{ u_{k-1}^i + {K_p} ( k_\alpha e^{-k_\beta k} - y_{k-1}) \right\}}_{u_{k}^i}}
     \end{array}
\end{equation}
where: $y^\ast$ is the output reference trajectory; $K_p$ and $K_I$ are real positive tuning gains; $\varepsilon_{k-1} = y^\ast_{k-1} - y_{k-1}$ is the tracking error; 
$k_\alpha e^{-k_\beta k}$ is an initialization function where $k_\alpha$ and $k_\beta$ are real constants;
practically, the integral part is discretized using e.g. Riemann sums. We define the set of $\mathcal{C}_{\pi}$-parameters of the controller as the set of coefficients 
$\{K_p, K_i, k_\alpha, k_\beta\}$. The internal recursion on $u_{k}^i$ is defined such as: $u_{k}^i = u_{k-1}^i + {K_p} ( k_\alpha e^{-k_\beta k} - y_{k-1})$. 

\section{Application to robot point-to-point movements}

\subsection{Controllable autonomous dynamical systems}

Robot discrete motions are modeled by autonomous Dynamical Systems (DS) that describe the behavior expected by the robot to perform tasks \cite{khansari}. 
Consider a state variable $\boldsymbol{\xi} \in \mathbb{R}^d$ that can be used to 
unambiguously define a discrete motion of a robotic system (e.g. $\xi$ could be a robot's joint angles,
position and/or orientation of an arm's end-effector in the operational space, etc.) and define the controllable function $\hat{f}$ such that:

\begin{equation}\label{eq:eq_kh}
 \boldsymbol{\dot{\xi}} = {f}( \boldsymbol{\xi} ) = \hat{f}( \boldsymbol{\xi} ) + \mathbf{u}, \qquad \mathbb{R}^d \mapsto \mathbb{R}^d
\end{equation}

where $\mathbf{u}$ is the input that allows controlling the model; $f(\boldsymbol{\xi})$ is a continuous function that codes an {\it exact} specific behavior of the robot 
and $\hat{f}(\boldsymbol{\xi})$ is the estimated function, derived from $f(\boldsymbol{\xi})$ 
\footnote{An estimate of $f(\boldsymbol{\xi})$ is built from a set of $N$ demonstrations using any of the state-of-the-art regression methods (see \cite{khansari}).} that needs
to be controlled in order to {\it comply} with the expected behavior. The expression (\ref{eq:eq_kh}) is integrated using an Euler forward method
\footnote{The following standard scheme is used: $\boldsymbol{\xi}_{k+1} = \boldsymbol{\xi}_k + h \boldsymbol{\dot{\xi}}$ (where $\boldsymbol{\dot{\xi}}$ is
deduced from the estimated function $\hat{f}(\boldsymbol{\xi})$) but when applying (\ref{eq:iPI_discret_nm_eq}) to close the 
loop (\ref{eq:iPI_discret_nm_eq_app}), 
little oscillations  of the trajectory (which remain to study / explain) appear but the closed-loop remains globally "dynamically" stable. To cancel these oscillations,
we notice that if one considers a $\mu$ factor such as: $\boldsymbol{\xi}_{k+1} = \mu \boldsymbol{\xi}_k + 
h \boldsymbol{\dot{\xi}}, \, \mu \in [0, 1]$, the modified Euler scheme allows having very nice dynamical performances in closed-loop despite an open-loop trajectory that does not correspond to the the original 
one (Fig. \ref{essai_0_khansari}) due to the presence of the "disturbing" $\mu$ factor inside the Euler scheme. }. 
Denote $\boldsymbol{\xi}_0 = \mathbf{0}$ the initial configuration and $\boldsymbol{\xi}_f$, the final point 
that must be reached by the controlled DS. 

\vspace{0.5cm}
{\it The purpose is to control (\ref{eq:eq_kh}) by the para-model law (\ref{eq:iPI_discret_nm_eq}) in order to maintain $\boldsymbol{\xi}$ "as close as possible" to a trajectory reference 
$\boldsymbol{\xi}^*$ according to the time.}

\subsection{Implementation of the $\mathcal{C}_{\pi}$-controller}

A possible control scheme is to consider controlling the trajectory $\boldsymbol{\xi}$ that must remain "as close as possible" to the reference $\boldsymbol{\xi}^*$. 
Therefore, $\boldsymbol{\xi}$ is physically measured and the position of the robot is driven by the $\mathcal{C}_{\pi}$-controller.
We build a closed-loop that creates a feedback between (\ref{eq:iPI_discret_nm_eq}) and (\ref{eq:eq_kh}). We have "symbolically":
\begin{equation} \label{eq:iPI_discret_nm_eq_app}
    \left\{  \begin{array}{c}
    \displaystyle{\mathbf{u}_k = \left. \int_0^t K_i (\boldsymbol{\xi}^*_{k-1} - \boldsymbol{\xi} _{k-1} ) d \, \tau \right|_{k-1} \underbrace{\left\{\mathbf{u}^i_{k-1} + {K_p} ( k_\alpha e^{-k_\beta k} - \boldsymbol{\xi}_{k-1}) \right\}}_{\mathbf{u}^i_{k}} } \\    
    \\[0.1cm]
    \boldsymbol{\dot{\xi}_k} =  \hat{f}( \boldsymbol{\xi}_k ) + \mathbf{u}_k
     \end{array} \right.
\end{equation}

\newpage
\subsection{Results}

\subsubsection{Lyapunov-based dynamical performances}

\begin{figure}[!h]
\centering
\includegraphics[width=12cm]{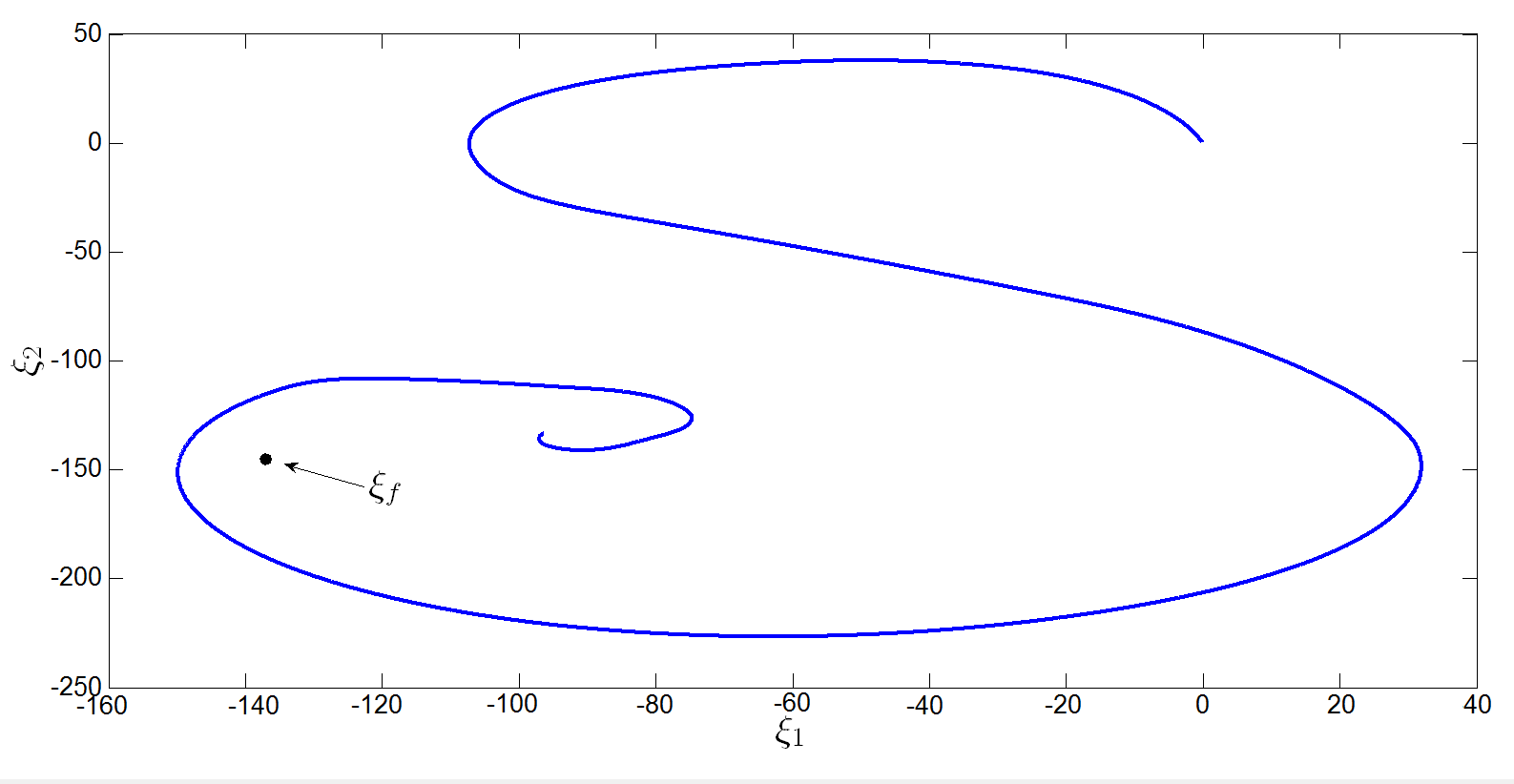}
\caption{Evolution of the trajectory $\boldsymbol{\xi}$ in open-loop ($\mathbf{u} = \mathbf{0}$).}
\label{essai_0_khansari}
\end{figure}

\begin{figure}[!h]
\centering
\includegraphics[width=12cm]{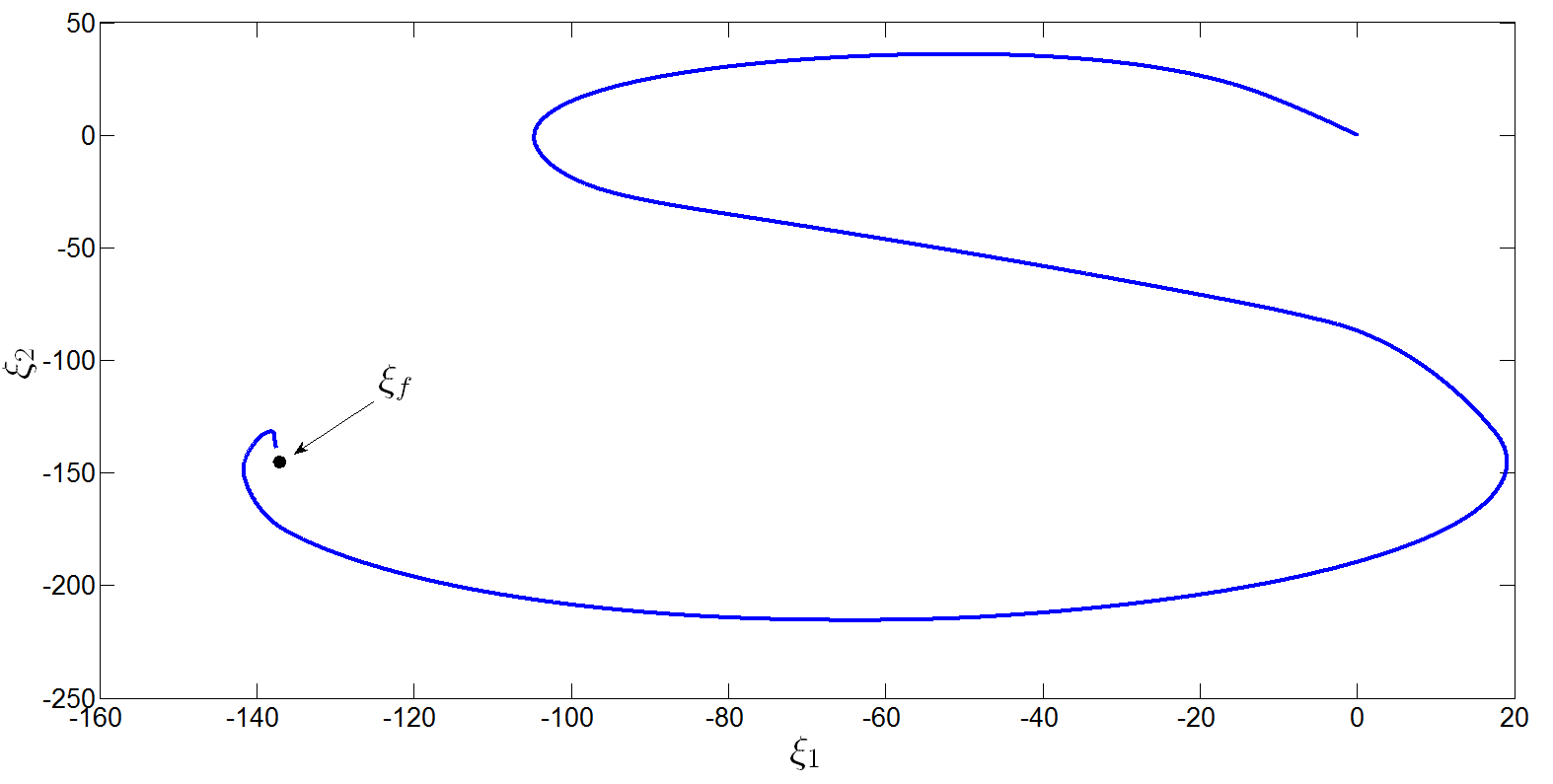}
\caption{Evolution of the trajectory $\boldsymbol{\xi}$ controlled with a Lyapunov approach.}
\label{essai_xf_khansari}
\end{figure}

Figure \ref{essai_0_khansari} presents the trajectory of the robot in open-loop i.e. described exclusively by $\hat{f}( \boldsymbol{\xi} )$; 
it shows that the state $\boldsymbol{\xi}$ converges to a point that is pretty far from the expected final point $\boldsymbol{\xi}_f$.
Figure \ref{essai_xf_khansari} presents the trajectory of the robot driven by the Lyapunov function approach \cite{khansari} that reaches the final point $\boldsymbol{\xi}_f$.


\begin{figure}[!h]
\centering
\includegraphics[width=12cm]{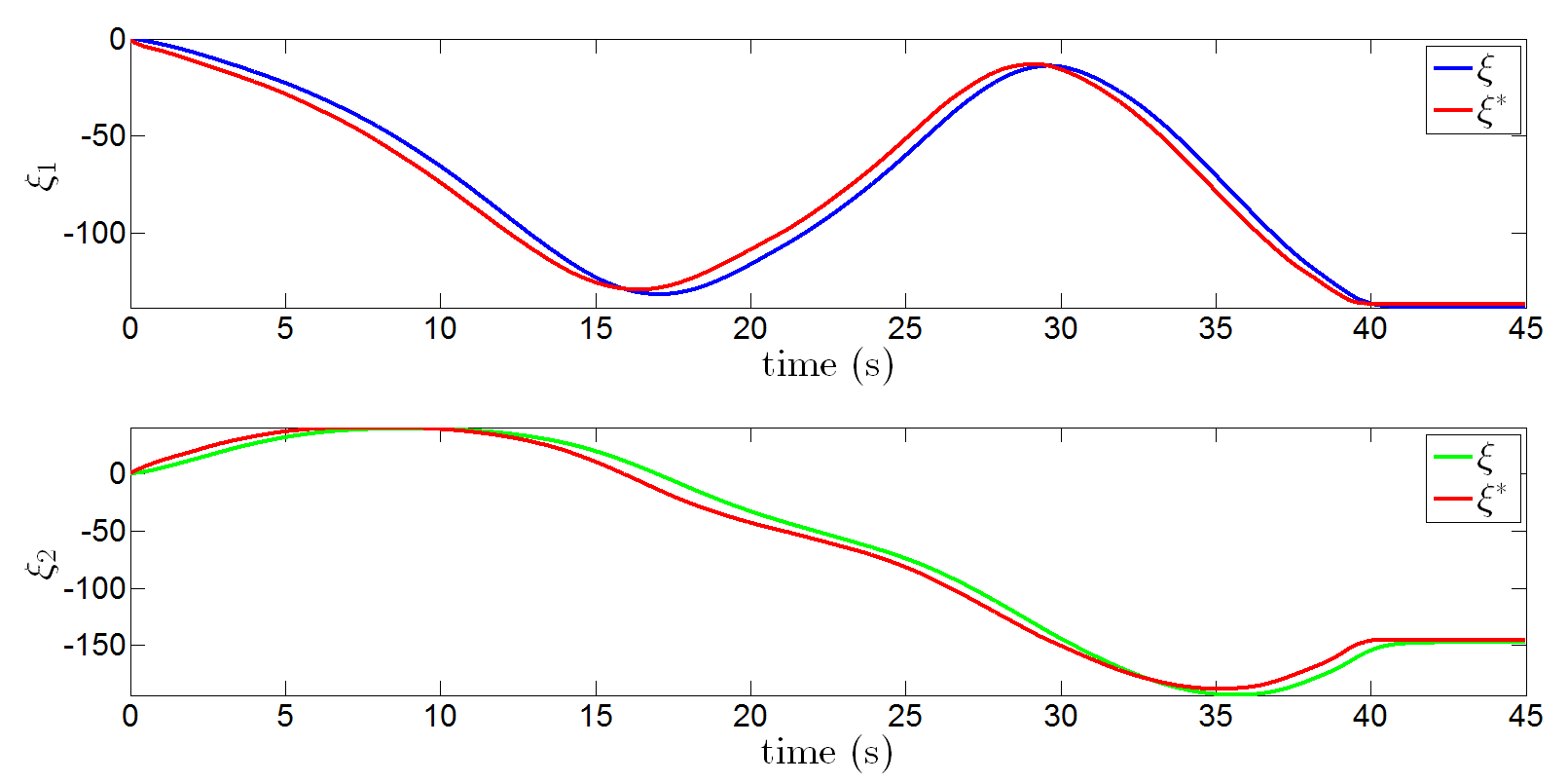}
\caption{Evolution of the trajectory $\boldsymbol{\xi}$ with respect to the reference $\boldsymbol{\xi}^*$ ($"1"$) according to the time.}
\label{essai_1_khansari}
\end{figure}

\begin{figure}[!h]
\centering
\includegraphics[width=12cm]{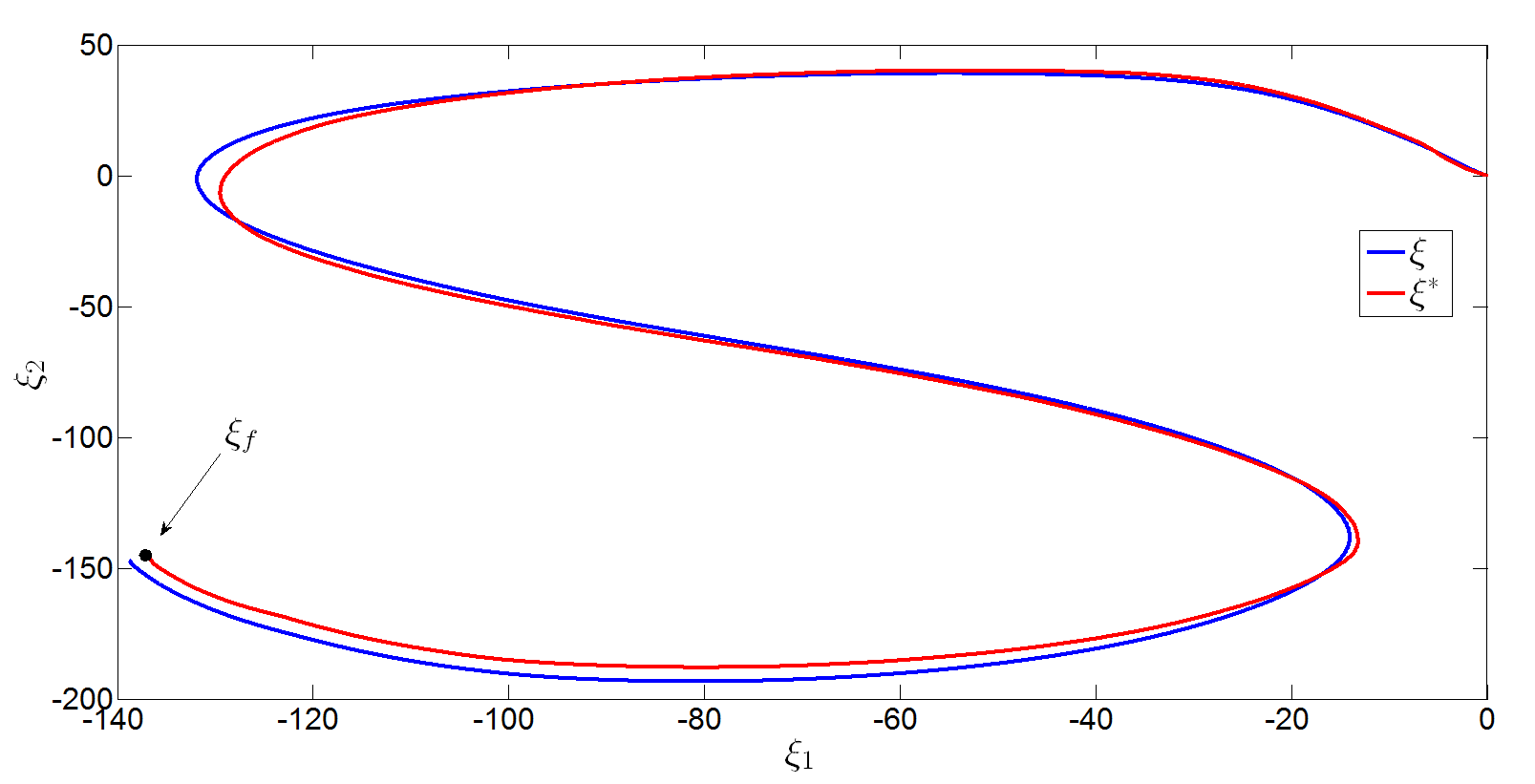}
\caption{Evolution of the trajectory $\boldsymbol{\xi}$ with respect to the reference $\boldsymbol{\xi}^*$ ($"1"$) in the phase space.}
\label{essai_12_khansari}
\end{figure}

Figure \ref{essai_1_khansari} shows the controlled trajectory $\boldsymbol{\xi}$ by the proposed para-model control according to the time for a particular reference $"1"$ and Fig. 
\ref{essai_12_khansari} shows the same result in the phase space.

\begin{figure}[!h]
\centering
\includegraphics[width=12cm]{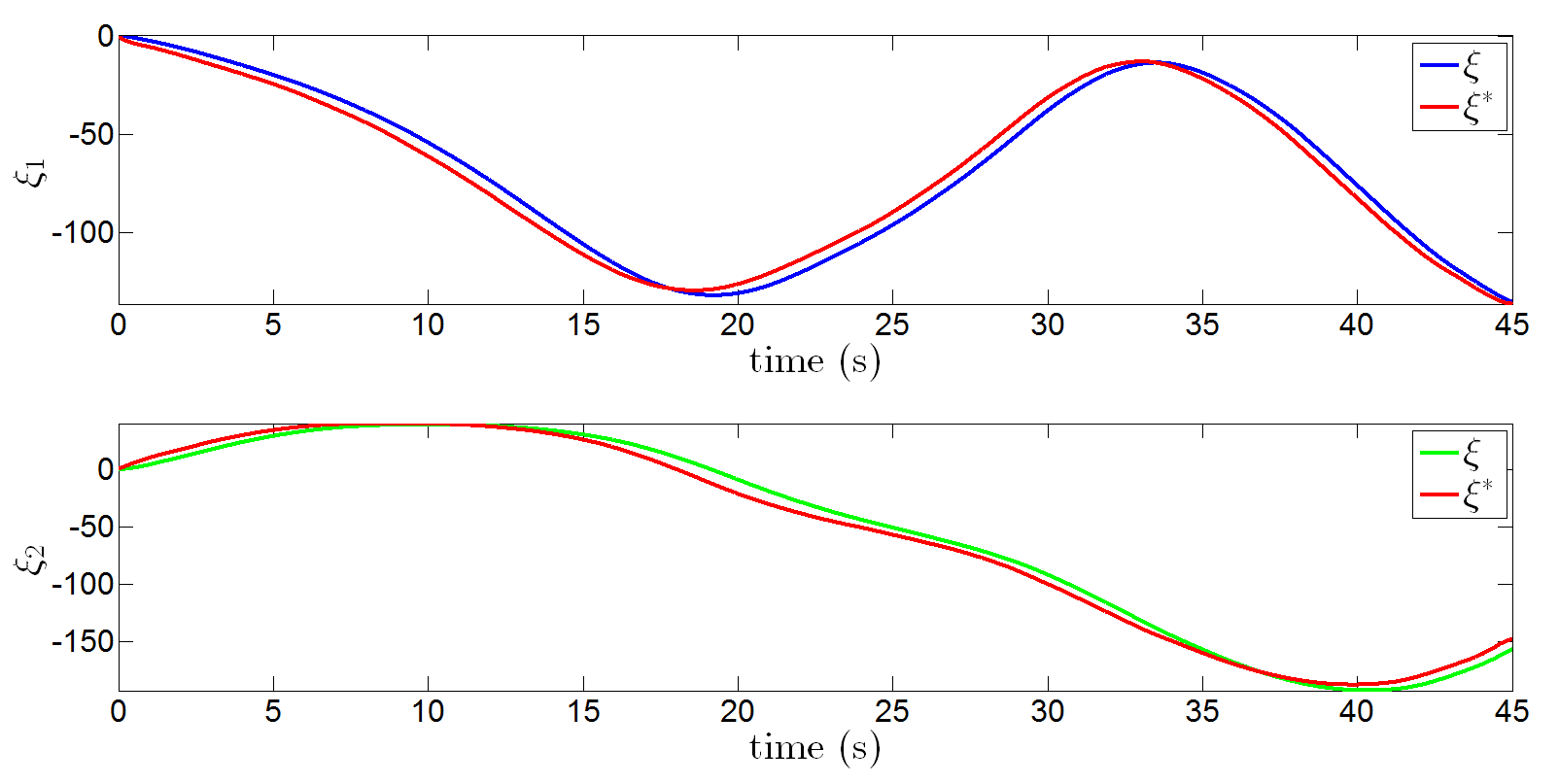}
\caption{Evolution of the trajectory $\boldsymbol{\xi}$ with respect to the reference $\boldsymbol{\xi}^*$ ($"2"$) according to the time.}
\label{essai_21_khansari}
\end{figure}

\begin{figure}[!h]
\centering
\includegraphics[width=12cm]{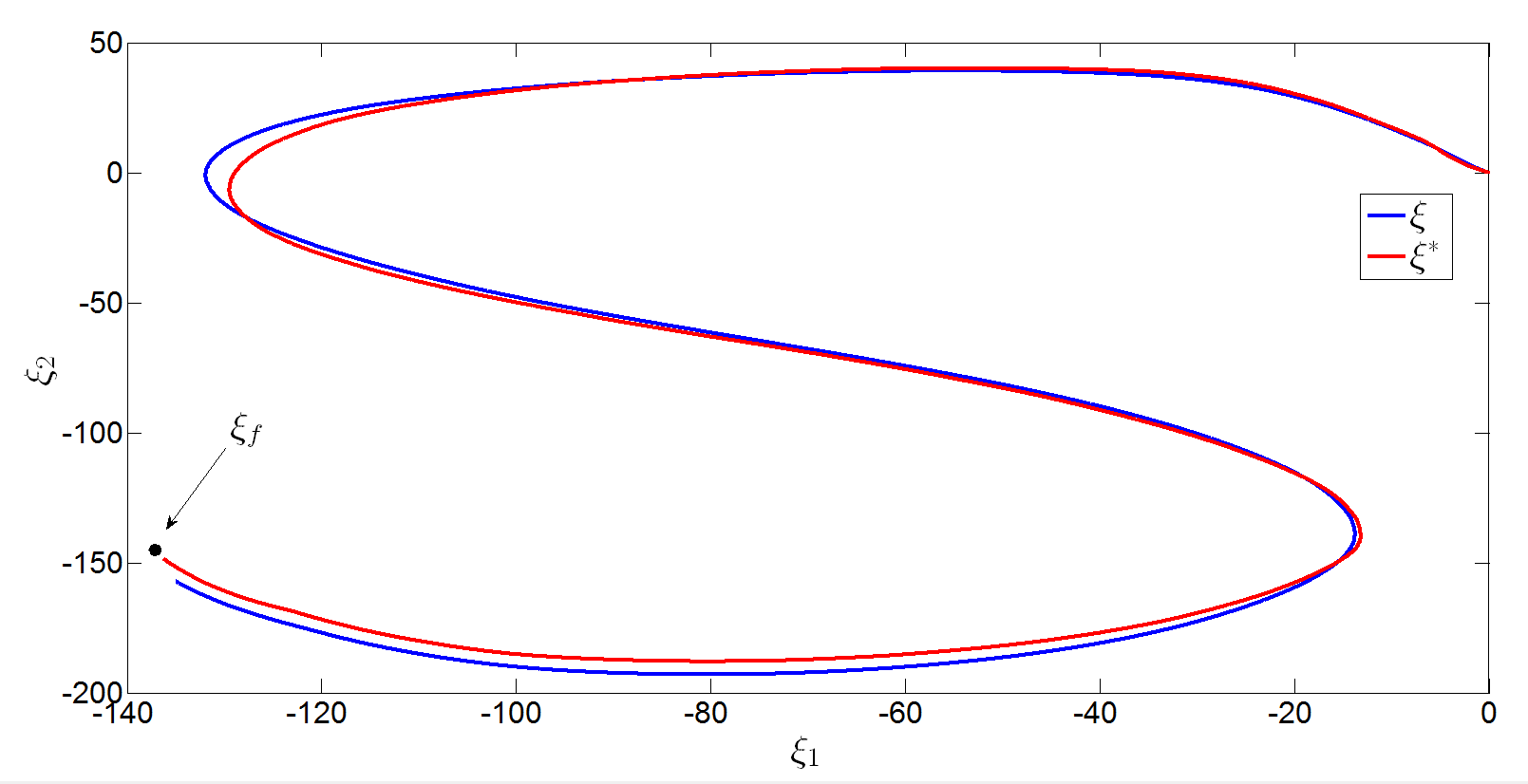}
\caption{Evolution of the trajectory $\boldsymbol{\xi}$ with respect to  the reference $\boldsymbol{\xi}^*$ ($"2"$) in the phase space.}
\label{essai_22_khansari}
\end{figure}

Figure \ref{essai_21_khansari} shows the controlled trajectory $\boldsymbol{\xi}$ by the proposed para-model control according to the time for a particular reference $"2"$ and Fig. 
\ref{essai_22_khansari} shows the same result in the phase space. According to the gained experience, the parameters $\{K_p, K_i, k_\alpha, k_\beta\}$ of the para-model law (\ref{eq:iPI_discret_nm_eq}) 
are very flexible and might give interesting dynamical performances in closed-loop even if they have been roughly tuned.

%\clearpage

\subsubsection{Optimized dynamical performances}

\begin{figure}[!h]
\centering
\includegraphics[width=12cm]{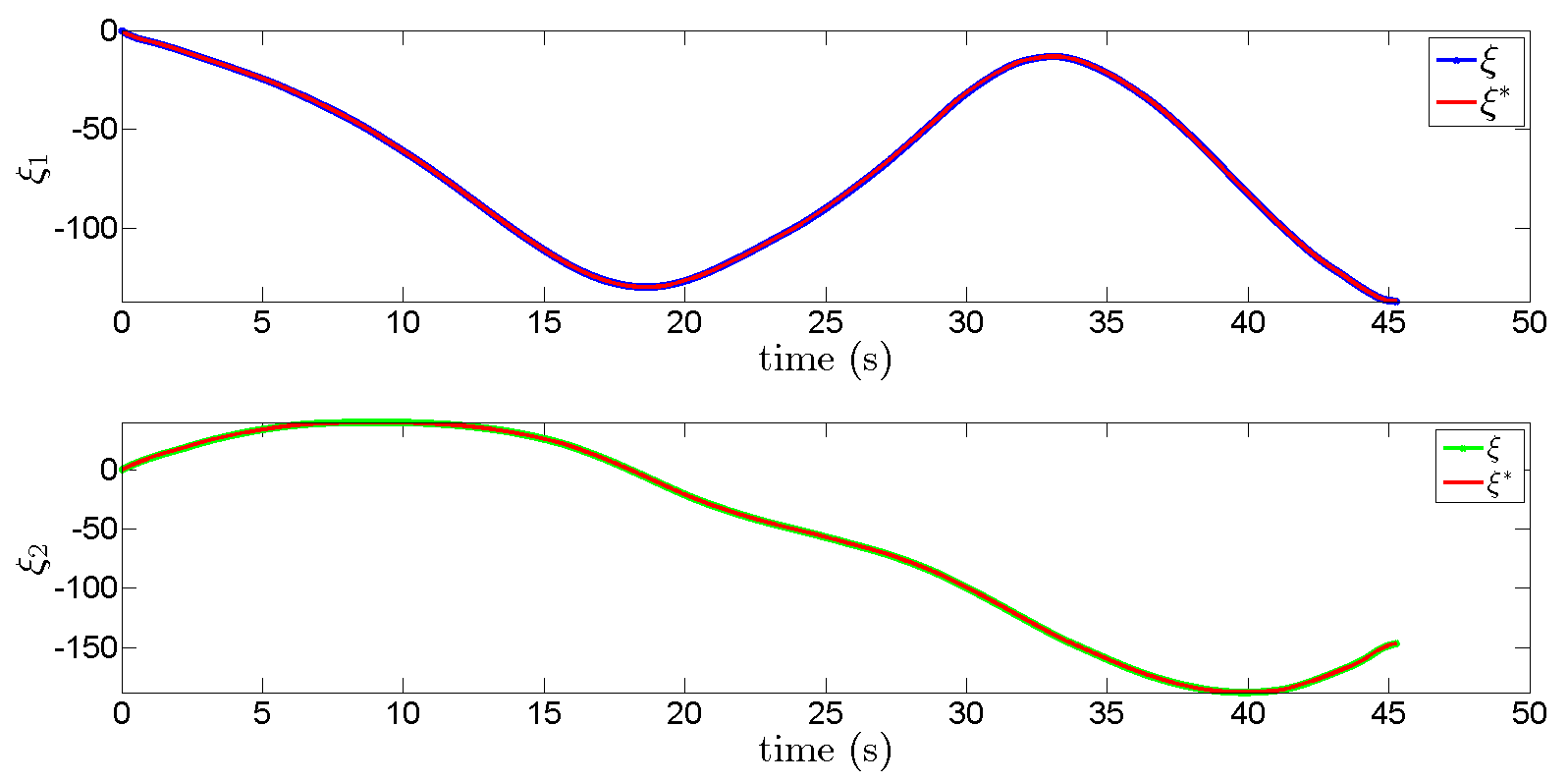}
\caption{Evolution of the optimized trajectory $\boldsymbol{\xi}$ with respect to the reference $\boldsymbol{\xi}^*$ ($"2"$) according to the time.}
\label{essai_31_khansari}
\end{figure}

\begin{figure}[!h]
\centering
\includegraphics[width=12cm]{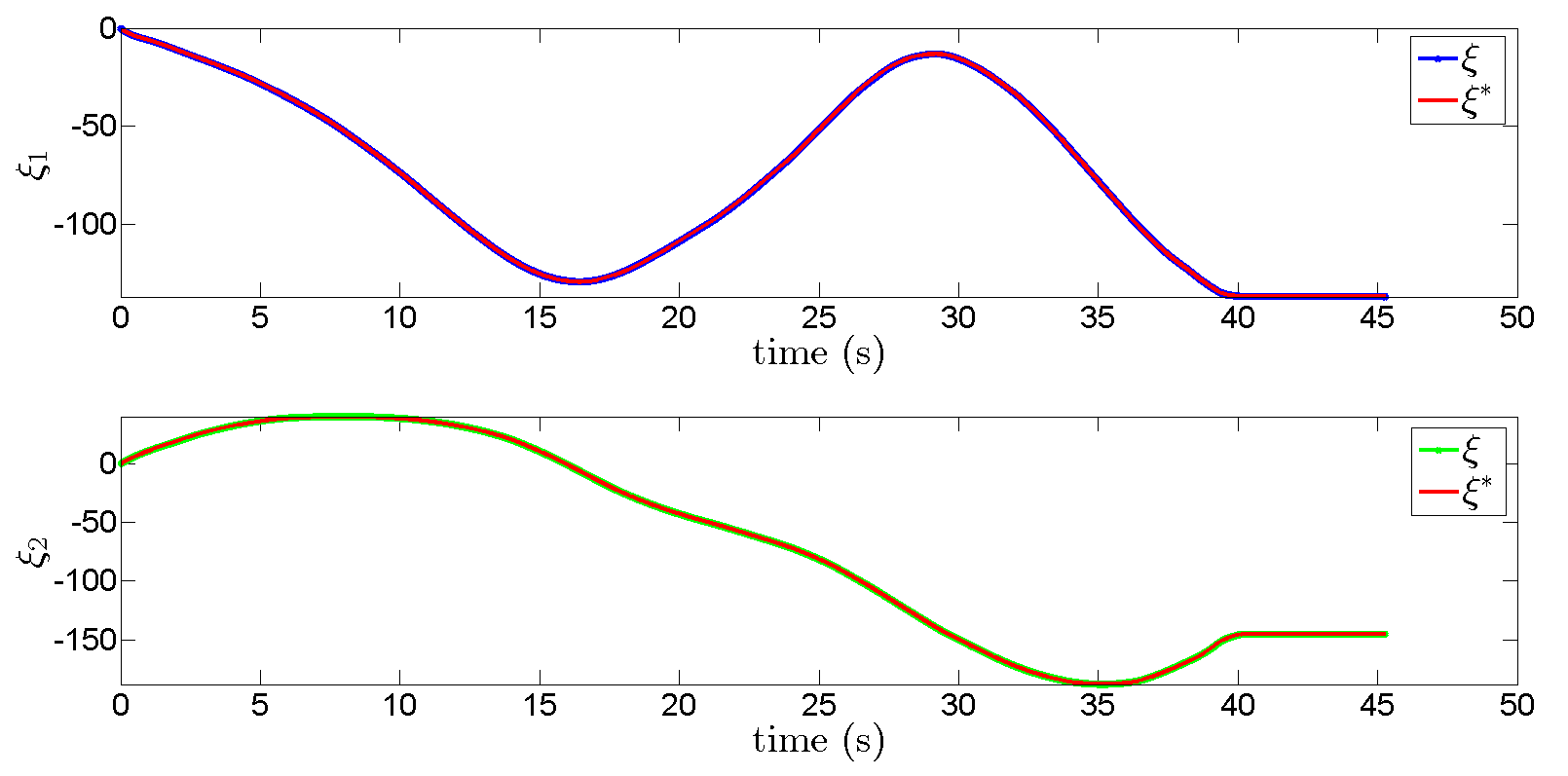}
\caption{Evolution of the optimized trajectory $\boldsymbol{\xi}$ with respect to the reference $\boldsymbol{\xi}^*$ ($"1"$) according to the time.}
\label{essai_41_khansari}
\end{figure}

To improve the dynamical performances of the closed-loop, we want to solve the problem of finding the most appropriate set of $\mathcal{C}_{\pi}$-parameters relating to the minimization of the ISE (integral square error)
performance index  such that:
\begin{equation*}
  \min_{K_p, K_i, k_\alpha, k_\beta} \int_0^{t_f} (\boldsymbol{\xi} - \boldsymbol{\xi}^*)^2 \diff \, t
 \end{equation*}
 \noindent
 where $t_f$ is the final time of the simulation. We are interested in using the "Brute Force Optimization" (BFO) solver \cite{BFO} that is very convenient and efficient to use. Figures 
\ref{essai_31_khansari} and \ref{essai_41_khansari} show the BFO-optimized controlled trajectory $\boldsymbol{\xi}$ by the proposed para-model control according to the time 
for respectively the references $"2"$ and $"1"$.
 
\subsection{Controlled trajectory with external disturbances}

To evaluate the disturbance rejection of the $\mathcal{C}_{\pi}$-controller, we consider adding an external "force" $\mathbf{u}^{dist}$ in (\ref{eq:eq_kh}) such as:

\begin{equation}\label{eq:eq_kh_dist}
 \boldsymbol{\dot{\xi}} = {f}( \boldsymbol{\xi} ) = \hat{f}( \boldsymbol{\xi} ) + \mathbf{u} + \mathbf{u}^{dist}
\end{equation}

\begin{figure}[!h]
\centering
\includegraphics[width=12cm]{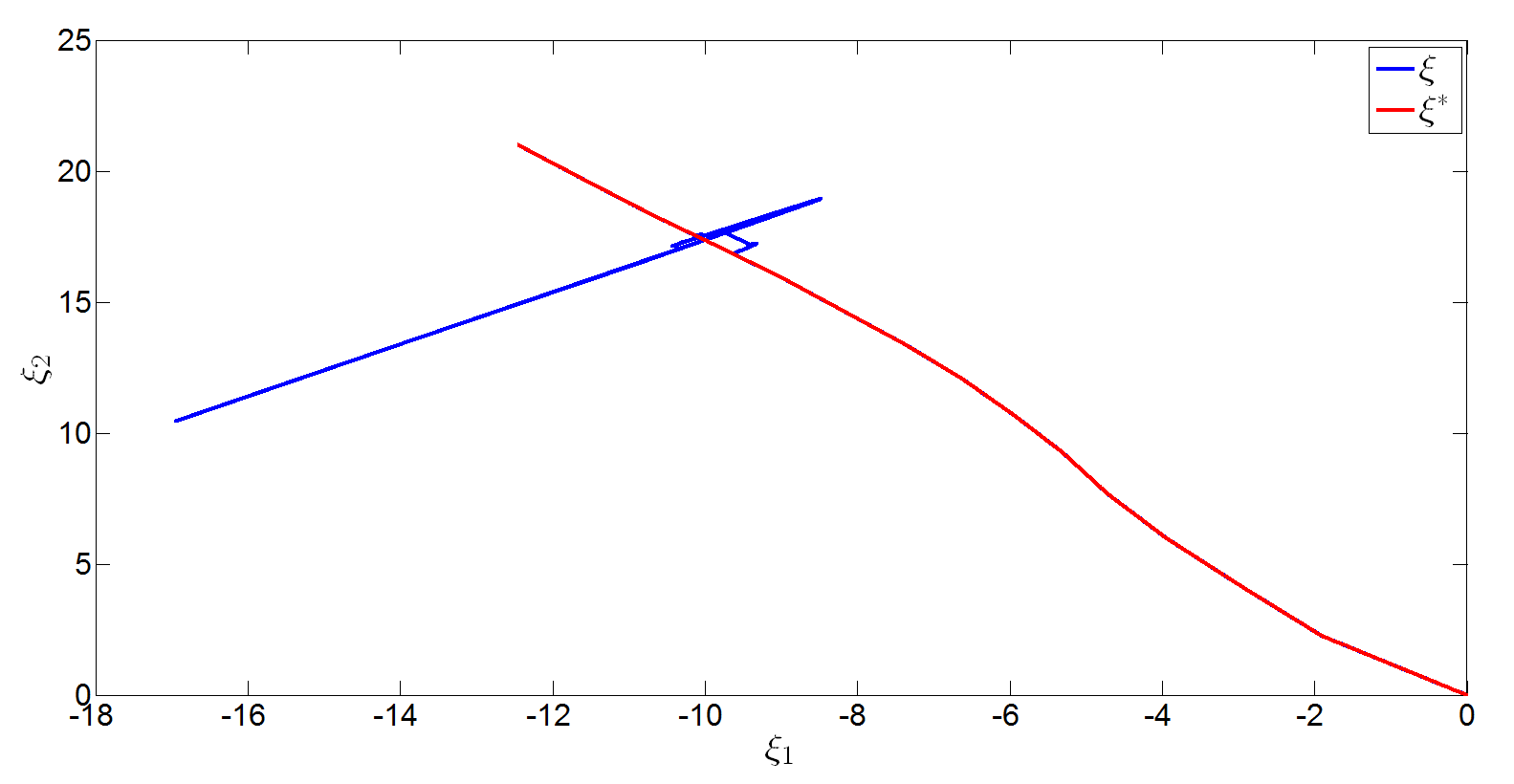}
\caption{Evolution of the disturbed controlled trajectory considering a linear disturbance (the disturbance is completely rejected after $t_{\beta}$).}
\label{essai_dist1_khansari}
\end{figure}

\begin{figure}[!h]
\centering
\includegraphics[width=12cm]{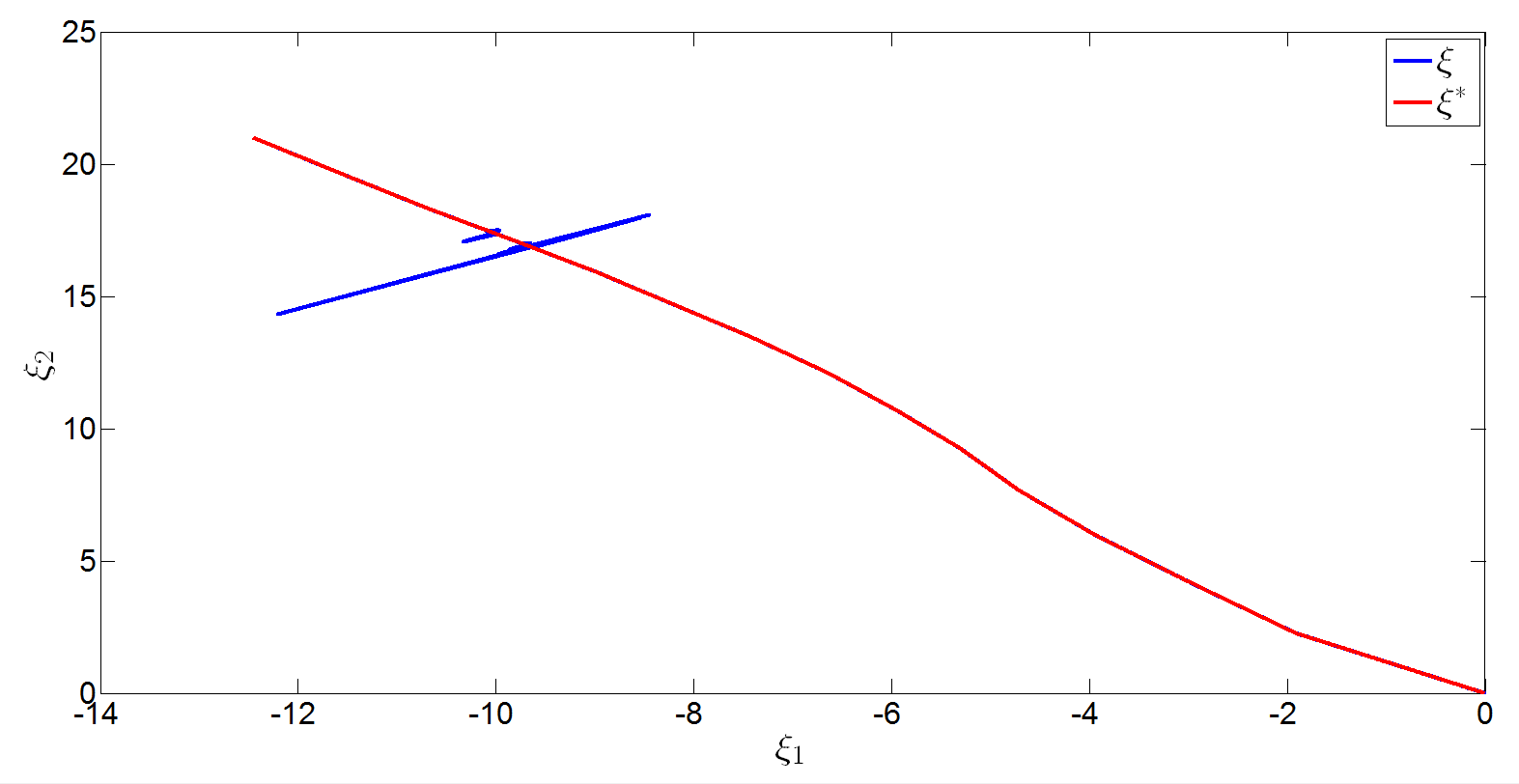}
\caption{Evolution of the disturbed controlled trajectory considering a logarithmic disturbance (the disturbance is completely rejected after $t_{\beta}$).}
\label{essai_dist2_khansari}
\end{figure}

The following examples illustrate the behavior of the controlled trajectory considering two cases of increasing disturbances: a linear-type disturbance (Fig. \ref{essai_dist1_khansari}) and a 
logarithmic-type disturbance (Fig. \ref{essai_dist2_khansari}).

\subsubsection{Examples}

We consider applying a disturbance $\mathbf{u}^{dist}_k$ over a small period $[t^{\alpha}, t^{\beta}] = [1.74, 1.81]$.

\paragraph{case 1 :}

\begin{equation}
\left\{ \begin{array}{l}
\mathbf{u}^{dist}_k = 0.1 + \mathbf{u}^{dist}_{k-1} \quad \text{when} \,\,  t_{\alpha} < t_k <t_{\beta} \\
\mathbf{u}^{dist}_k = 0 \quad \text{when} \,\, t_k < t_{\alpha} \,\, \text{and}  \,\, t_k > t_{\beta}
\end{array} \right. \quad \text{with} \,\, \mathbf{u}^{dist}_0 = 0.1
\end{equation}

\paragraph{case 2 :}

\begin{equation}
\left\{ \begin{array}{l}
\mathbf{u}^{dist}_k = \ln(\mathbf{u}^{dist}_{k-1}) \quad \text{when} \,\, t_{\alpha} < t_k <t_{\beta} \\
\mathbf{u}^{dist}_k = 0 \quad \text{when} \,\, t_k < t_{\alpha} \,\, \text{and}  \,\, t_k > t_{\beta}
\end{array} \right. \quad \text{with} \,\, \mathbf{u}^{dist}_0 = 1.1
\end{equation}

\section*{Acknowledgement}

The author is sincerely grateful to Dr. Edouard Thomas for his strong guidance and his valuable comments that improved this paper.


\begin{thebibliography}{}

\bibitem{fliess2} M. Fliess and C. Join, "Model-free control", International Journal of Control, vol. 86, issue 12, pp. 2228-2252, Jul. 2013 (available at \url{http://arxiv.org/pdf/1305.7085.pdf}).

\bibitem{kumar} K.J. \r{A}str\"{o}m and P.R. Kumar, "Control: A perspective", Automatica, vol. 50, issue 1, pp. 3-43, Jan. 2014 (available at \url{http://www.sciencedirect.com/science/article/pii/S0005109813005037}).

\bibitem{michel} L. Michel, "A para-model agent for dynamical systems", preprint arXiv, Oct. 2016 (available at \url{http://arxiv.org/abs/1202.4707}).

\bibitem{khansari} S.M. Khansari-Zadeh and A. Billard, "Learning Control Lyapunov Function to Ensure Stability of Dynamical System-based Robot Reaching Motions". Robotics and Autonomous Systems, 
vol. 62, no 6, pp. 752-765, 2014 (available at \url{http://lasa.epfl.ch/publications/uploadedFiles/Khansari_Billard_RAS2014.pdf}).

\bibitem{BFO} M. Porcelli and Ph. L. Toint, BFO, "A trainable derivative-free Brute Force Optimizer for nonlinear bound-constrained optimization and equilibrium computations with continuous and discrete variables",
\textit{Namur Center for Complex Systems}, In Support, Tech. Reports, no. naXys-06-2015, Belgium, Jul. 2015 (available at \url{http://www.optimization-online.org/DB_FILE/2015/07/4986.pdf}).

%\bibitem{Tarraf} D.C. Tarraf, A. Megretski and M.A. Dahleh,  "A Framework for Robust Stability of Systems Over Finite Alphabets," IEEE Transactions on Automatic Control, vol.53, no.5, pp.1133-1146, June 2008.

%\bibitem{Tarraf2} D.C. Tarraf, A. Megretski and M.A. Dahleh, "Finite Approximations of Switched Homogeneous Systems for Controller Synthesis," IEEE Transactions on Automatic Control,  vol.56, no.5, pp.1140,1145, May 2011.

%\bibitem{Tarraf3} D.C. Tarraf, "On a control-oriented notion of finite state approximation," in IEEE International Symposium on  Computer-Aided Control System Design (CACSD), pp.1385,1390, 28-30 Sept. 2011.

%\bibitem{Tarraf4} D.C. Tarraf, "A Control-Oriented Notion of Finite State Approximation," IEEE Transactions on Automatic Control, vol.57, no.12, pp.3197,3202, Dec. 2012.

\end{thebibliography}
\end{document}